\documentclass[numberedappendix]{emulateapj}

\usepackage{amsmath} 
\usepackage{subfigure}
\usepackage[]{graphicx}
\usepackage{multirow}

\def\Fe{$^{55}${Fe}~}

\begin{document}


\title{The \Fe X-ray Energy Response of Mercury Cadmium Telluride Near-Infrared Detector Arrays}

\author{Ori D. Fox\altaffilmark{1,2}, Augustyn Waczynski\altaffilmark{3}, Yiting Wen\altaffilmark{4}, Roger D. Foltz\altaffilmark{5}, Robert J. Hill\altaffilmark{6}, Randy A.  Kimble\altaffilmark{2}, Eliot Malumuth\altaffilmark{7}, and Bernard J. Rauscher\altaffilmark{2}}
\altaffiltext{1}{Department of Astronomy, University of Virginia, Charlottesville, VA 22903}
\altaffiltext{2}{NASA Goddard Space Flight Center, Greenbelt, MD 20771}
\altaffiltext{3}{Global Science and Technologies, Inc., Greenbelt, MD 20771}
\altaffiltext{4}{Muniz Engineering, Inc., Lanham, MD 20706}
\altaffiltext{5}{Sigma Space, Inc., Lanham, MD 20706}
\altaffiltext{6}{Conceptual Analytics, LLC, Glenn Dale, MD 20769}
\altaffiltext{7}{Wyle Information Systems, McLean, VA 22102}
\email{ofox@virginia.edu}

\begin{abstract}

A technique involving \Fe X-rays provides a straightforward method to measure the response of a detector.  The detector's response can lead directly to a calculation of the conversion gain ($e^-~{\rm ADU}^{-1}$), as well as aid detector design and performance studies.  We calibrate the \Fe X-ray energy response and pair production energy of HgCdTe using 8 HST WFC3 1.7 \micron~flight grade detectors.  The results show that each K$\alpha$ X-ray generates 2273~$\pm$~137 electrons, which corresponds to a pair-production energy of  2.61~$\pm$~0.16 eV.  The uncertainties are dominated by our knowledge of the conversion gain.  In future studies, we plan to eliminate this uncertainty by directly measuring conversion gain at very low light levels.

\end{abstract}


\section{Introduction}
\label{sec:intro}

Conversion gain, $g_c$ ($e^-~{\rm ADU}^{-1}$), is a fundamental parameter in detector characterization that is used to measure many detector properties, including read noise, dark current, and quantum efficiency (QE).  Measuring these parameters to a higher precision is becoming more important as the demand for low-signal observations increases and the scientific requirements evolve in complexity.  

The upcoming Joint Dark Energy Mission (JDEM), for example, requires the precise measure of the dark energy equation equation of state, $w$ and $w'$.  This measurement is derived from the luminosity distance versus redshift relationship of Type Ia supernovae out to high redshifts. To obtain the necessary accuracy, the mission requires detector characterization measurements to within 2\% \citep{sch:06}.

\begin{figure} 
\begin{center}
\subfigure[Symmetric PSF]{
\label{f1a}
\epsscale{0.75}
\plotone{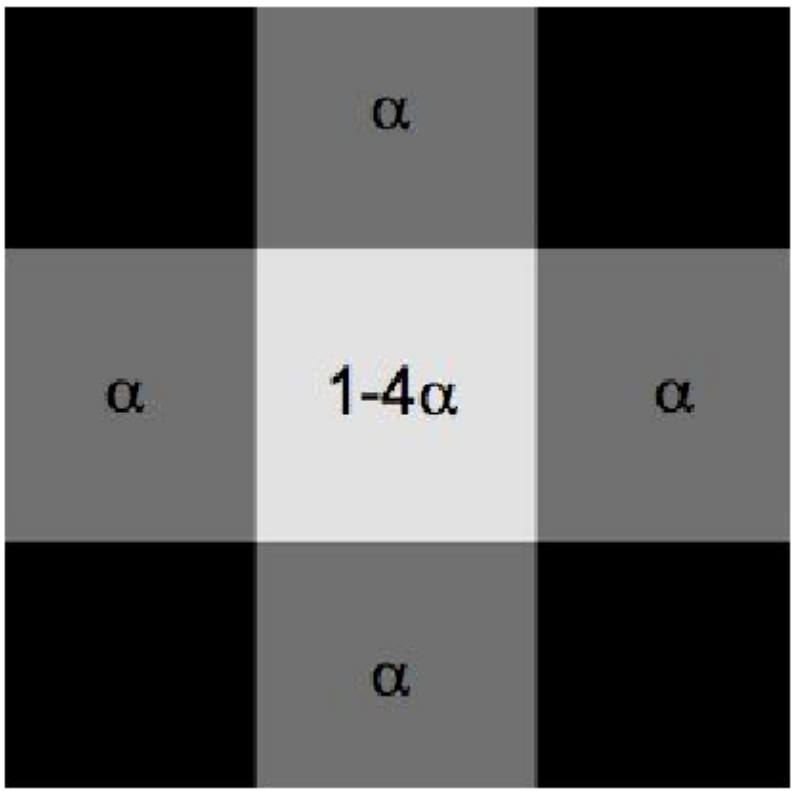}}\\
\subfigure[Partially Symmetric PSF]{
\label{f1b}
\epsscale{0.75}
\plotone{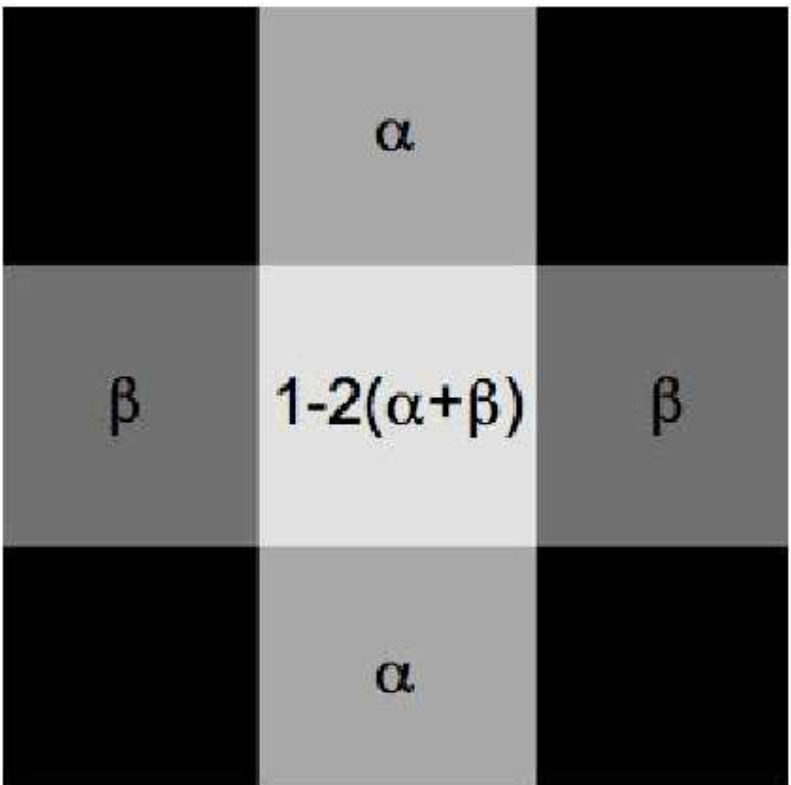}}
\caption{Illustration of charge coupling mechanism for both the (a) symmetric and (b) partially-symmetric cases.  In both cases, photocurrent entering a central pixel may be displaced into neighboring pixels by coupling mechanisms such as interpixel capacitance (IPC).  The fraction of charge coupled into the neighboring pixels is given by the coupling coefficients, $\alpha$ and $\beta$.}
\label{f1}
\end{center}
\end{figure}

A traditional method for measuring conversion gain is the photon transfer technique \citep{jan:87}.  This method, however, is limited in that it does not account for charge coupling between pixels.  Charge coupling attenuates the photon shot noise of individual pixels, resulting in an underestimated variance and an overestimated gain.  Figure \ref{f1} illustrates the coupling of the central pixel with the four nearest neighbors, both for (a) symmetric and (b) partially-symmetric cases.  The degree of coupling is given by coupling coefficients, $\alpha$ and $\beta$.

Interpixel capacitance (IPC), one of the more dominant charge coupling mechanisms, is caused by small amounts of stray capacitance.  Early simulations first predicted the effects of IPC \citep{kav:94}.  Later studies measured the degree of coupling using various methods, including cosmic ray hits \citep{fig:04}, externally calibrated capacitors \citep{fin:06}, and hot pixels \citep{bai:07}.  \citet{moo:06} provide a detailed mathematical discussion.  These works all indicate that ignoring IPC effect within infrared detectors can result in an overestimate of gain by up to 20\%.  \citet{dor:06} show these effects to be as a high as a factor of 2 within the fully depleted silicon-based hybrid CMOS arrays, such as the Rockwell HyViSI H2RG arrays.

While the conversion gain is ultimately measured via standard stars once the instrument has been launched and commissioned \citep[e.g.][]{reach:05}, detectors must meet specifications and science missions must be planned much earlier.  An overestimate of the gain can significantly distort the assessment of the detector and mission potential.  IPC becomes more noticeable in near-infrared detectors as pixel size decreases.  Better methods of measuring $g_c$ therefore have broad applicability to astronomical detector characterization.  For example, one popular method relates the strength of the inter-pixel coupling and, subsequently, the ``true'' variance to the autocorrelation function of the detector \citep{moo:06}.  We discuss this method further in Section \ref{sec:validation}.

A technique involving \Fe X-rays provides an alternative, straightforward method to measure the conversion gain.  This technique quantifies the necessary energy to liberate a single electron-hole pair in the semiconductor, known as the pair creation energy.  If the X-ray energy is known, the number of electrons liberated by each X-ray photon follows.  The conversion gain is then derived from the observed instrumental counts (ADU) produced by a single X-ray.  Already, this is a popular technique within the CCD community, where each \Fe~K$\alpha$ X-ray photon is known to liberate $\sim$1620 electrons in silicon \citep{fra:94}.

This absolute calibration offers additional benefits that aid detector design and performance studies.  For example, the calibrated response provides a fundamental standard for measuring drifts and/or changes in a detector's system gain over time.  Additionally, if the conversion gain is given by a reliable alternative method, such as \citet{moo:06}'s autocorrelation technique, the \Fe X-ray response serves as a direct probe of the internal quantum efficiency (QE), also known as the charge collection efficiency.  In other words, this technique measures the detector response to individual X-rays so that the QE does not include any contribution from external factors such as the anti-reflective coating or long-wavelength photons that are not absorbed.  While not commonly used by the astronomical community, the internal QE is an important measurement to detector manufacturers in their quest to build more efficient detectors.

A semi-empirical trend exists between the pair creation energy and semiconductor bandgap \citep{ali:75}:
\begin{equation}
\epsilon = 2.73E_g + 0.55,
\label{eqn:ali}
\end{equation}
where $\epsilon$ is the pair-production energy (in eV) and $E_g$ is the band-gap energy (in eV).  An exact pair-production energy, however, is not known for the commonly used mercury-cadmium-telluride (HgCdTe) infrared detector arrays.  The limiting factor has been the cadmium-zinc-tellruide (CdZnTe) substrate found on many HgCdTe detectors, which absorbs the \Fe X-rays before they reach the depletion region of the detector. The infrared detectors for the HST Wide-Field Camera 3 (WFC3) are not limited in this respect because they are all substrate-removed.

In this paper, we provide calibrated measurements of the \Fe X-ray response of 1.7 \micron~HgCdTe detectors.  Section~\ref{sec:methods} presents our \Fe data.  To convert the X-ray intensity from ADU to electrons, we establish the ``true'' variance and gain of a pixel using classical propagation of errors.  Section~\ref{sec:math} discusses the mathematical framework for this method.  Monte Carlo simulations validate the fidelity of this technique.  Section~\ref{sec:results} presents the results.   We find that in response to the \Fe X-rays, these 1.7 \micron~HgCdTe detectors yield 2273~$\pm$~137 electrons, which corresponds to a pair-production energy 2.61~$\pm$~0.16 eV.  We describe in detail the various systematic errors inherent within this measurement.

\section{Test Methodology and Data}
\label{sec:methods}

We have collected \Fe X-ray data from eight Teledyne Imaging System's (TIS) 1.7 \micron~H1R flight grade WFC3 detectors, which are molecular beam epitaxy (MBE) grown and CdZnTe substrate-removed.  These detectors have a read noise of $\sim$15 $e^-$~rms per read ($\sim$21 $e^-$ CDS) and a dark current of $\sim$0.02 $e^-~{\rm s}^{-1}$ at 145 K.  All data were collected at the NASA Goddard Space Flight Center (GSFC) Detector Characterization Laboratory (DCL).  The \Fe X-ray source was placed at a normal incidence 1.5 cm from the center of the 1.8$\times$1.8 cm FPA.  The data were all collected in 2008 and have since been archived.

For each detector, we obtained 10 exposures, each consisting of 6 correlated double samples (CDS) taken up-the-ramp, where each CDS pair is composed of the $0^{th}$~and $n^{th}$ frames up the ramp, where $n$~is a number between 1 and 6.  We performed a standard dark subtraction with median frames created from the 10 exposures.  We include plots for only FPA 152, which is a typical representation of the other detectors we tested.  We discuss any differences observed in the other detectors.

\begin{figure}[]
\begin{center}
\subfigure[Single Pixel Flux]{
\label{f2a}
\plotone{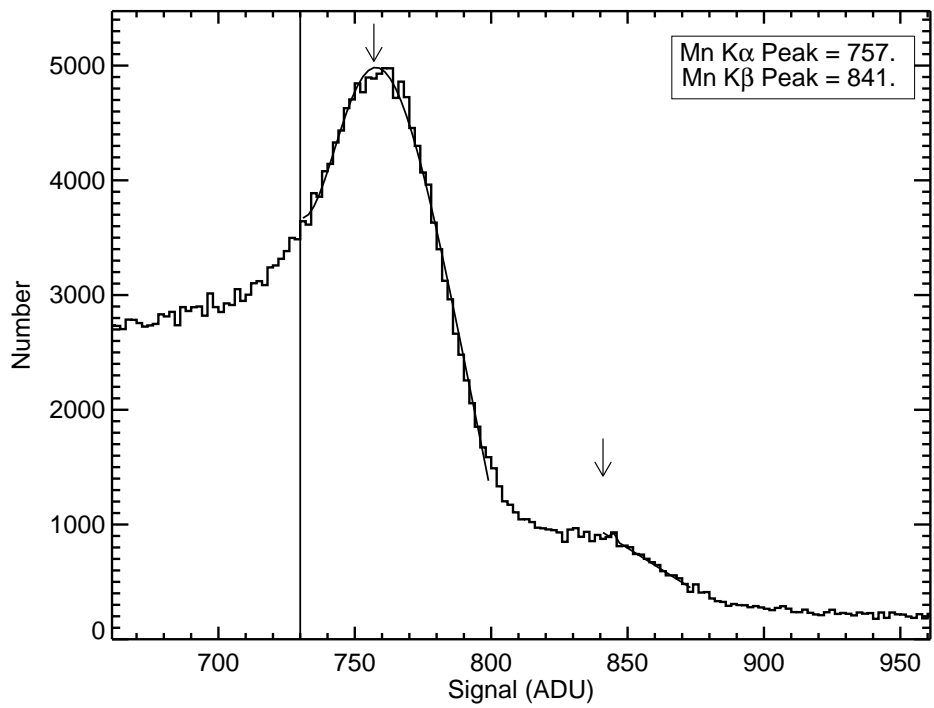}}\\
\subfigure[Nearest Neighbor Summation]{
\label{f2b}
\plotone{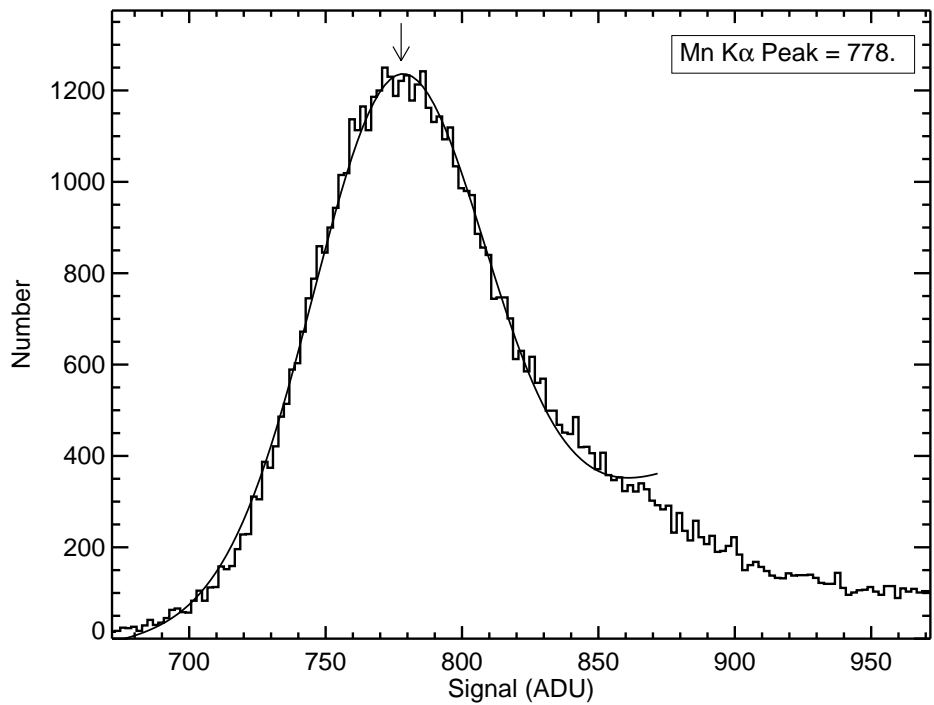}}
\caption{Histograms of the observed \Fe X-ray response, in ADU, for a typical 1.7 \micron~HgCdTe detector (FPA 152).  Figure \ref{f2a} shows only the individual pixel response, while Figure \ref{f2b} includes the summation of the nearest neighbor intensities.  The K$\alpha$ intensity peak is sharper and better defined by the individual pixels, but the histogram ignores charge coupling effects.  Adding the nearest neighbor pixel intensities compensates for charge coupling and thereby provides a more accurate measurement of the true X-ray response.}
\label{f2}
\end{center}
\end{figure}

Figure \ref{f2a} shows a histogram of individual pixel intensities (in ADU) from the 10 exposures of FPA 152.  Key features include the dominant K$\alpha$ peak, the less dominant K$\beta$ peak, and a low-intensity plateau (that extends down to zero signal).  The two peaks show broadening likely due to a combination of several effects.  Off-centered X-ray hits and charge diffusion, which occurs on the 0.1-pixel level, cause broadening to the left.  A changing band-gap energy with penetration depth causes broadening to the right.  Read noise, pixel-to-pixel gain variations, and the intrinsic fano factor\footnote{The Fano Factor adjusts for additional line broadening due to the statistical nature of the loss of energy to thermal excitation in the lattice.  An accurate measure of the Fano factor requires detailed modeling of proprietary data from TIS.} cause overall broadening.  We fit the peak of each profile using IDL's GAUSSIAN procedure, which performs a combined gaussian and polynomial fit.  The low-intensity plateau results mostly from charge diffusion and non-centered hits.

\begin{figure}
\begin{center}
\epsscale{1}
\plotone{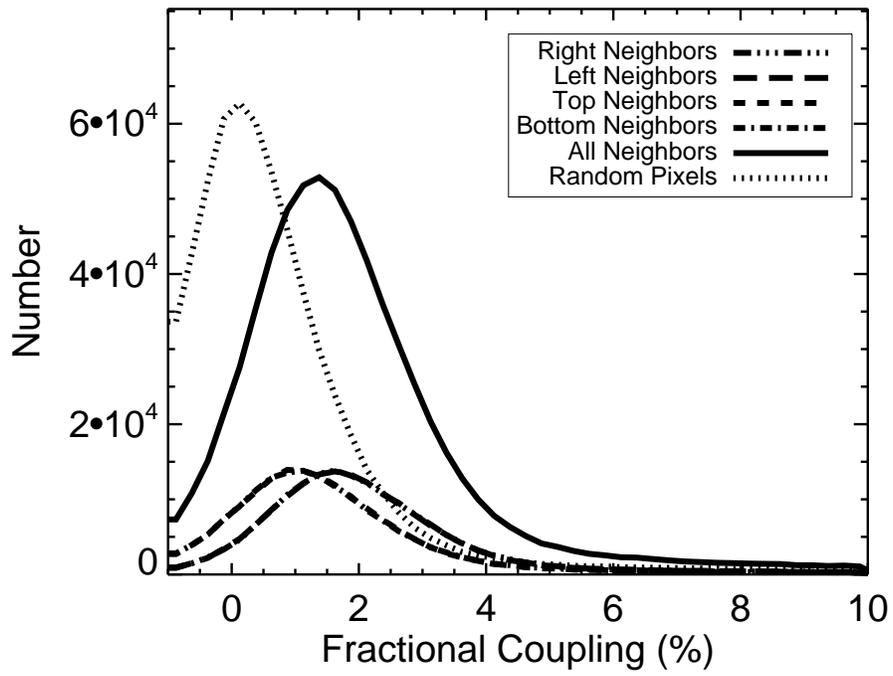}
\caption{Histogram of fractional crosstalk observed in neighboring pixels when compared to the total X-ray intensity of a typical detector (FPA 152).  The nearest neighbors show a significant degree (1-2\%) of fractional crosstalk with the central pixel.  In contrast, a distribution of random pixels shows no coupling the X-rays.}
\label{f3}
\end{center}
\end{figure}

The single pixel histogram in Figure \ref{f2a}, however, ignores charge coupling effects that cause some of the signal from the X-ray to appear in neighboring pixels, such as in Figure \ref{f1}.  The histogram in Figure \ref{f3} shows the fraction of the total charge in FPA 152 that couples to neighboring pixels (where the total charge is defined as the intensity in the central pixel and four nearest neighbors).  The plot reveals a 1-2\% coupling in all primary directions.  While there is a distinct symmetry in both the horizontal and vertical directions, the horizontal coupling tends to be larger than the vertical coupling.  This partial symmetry suggests a non-symmetric multiplexer associated with the horizontal readout mechanism.  Overall, all nearest neighbors (solid line) show an average coupling of $\sim$1.5\% while a distribution of random pixels (dotted line) shows, as expected, no coupling to the X-rays.

Adding the nearest neighbor flux presents the most straightforward method by which to restore the lost charge.  We do not perform aperture photometry.  While coupling is observed in other pixels, such as diagonal and second-neighbor pixels, Monte Carlo simulations reveal this effect to be negligible on the 1\% level.  The additional read noise from these pixels is greater than the actual signal.  We conclude that the additional noise from these pixels would be more detrimental to the total signal than the small amount of charge lost when adding only nearest neighbors.  We consider only hits that are isolated (defined within a 5$\times$5-pixel area) and well-centered (defined as being associated with the peak intensity in Figure \ref{f2a}, as opposed to the low-intensity plateau).  This threshold is indicated by the vertical line in Figure \ref{f2a}.  In this case, ``well-centered'' is a somewhat arbitrary definition, and we find that shifting the threshold can alter the results by tens of ADU.  We discuss the implications of this effect in our discussion of systematic errors in Section \ref{sec:syserr}.

Figure \ref{f2b} plots a histogram of only the selected pixel intensities with the nearest neighbor charge restored.  Expectedly, the peak of this distribution shifts to a higher intensity, consistent with the fractional coupling observed in Figure \ref{f3}.  The additional read noise from the neighboring pixels blends the K$\alpha$ and K$\beta$ peaks.  We again take the peak by fitting the distribution with IDL's GAUSSIAN routine.  The peak in this case represents the detector's response (in ADU) to a single \Fe X-ray, for which the energy now includes a weighted contribution from the K$\beta$ line. Given a K$\alpha$-to-K$\beta$ peak ratio of 7:1\citep{sco:74}, we approximate the X-ray energy as a weighted average: 5.9375 keV.

\section{Calculating Gain via Propagation of Errors}
\label{sec:math}

The pair-production energy of the detectors ultimately requires the total intensity of the X-rays to be in electrons.  The conversion of the detector response in Figure \ref{f2b} from ADU into electrons, therefore, must be calculated with an independent measure of the conversion gain.  The photon transfer method is insufficient because charge coupling attenuates the photon shot noise and decreases the measured variance.  Here, we explore a technique to establish the ``true'' variance via the classical propagation of errors.  The technique has the advantage of allowing for a direct measurement of the covariance matrix from the data, which can serve as a diagnostic of other correlations that may be present in addition to IPC.  We find this method works well at all signal levels and is consistent with the results of \citet{moo:06}.  We therefore implement this method throughout our analysis.

\subsection{Mathematical Preliminaries}
\label{sec:fullsym}

We begin by considering the simplest possible example of small symmetric crosstalk to the four nearest neighbors.  (A partially symmetric crosstalk consistent with Figure \ref{f3} is more complicated and considered in Section \ref{sec:partsym}.)  As we discuss in Section \ref{sec:methods}, we ignore diagonal and second-neighbor pixels because coupling effects are negligible on the 1\% level for our detectors.  For larger IPC, other neighbors must be considered.  

The detector is uniformly illuminated and only Poisson noise is present (i.e. no read noise).  This model is true for high signal levels for which shot noise in the signal dominates.  Crosstalk is assumed to be symmetric and to only affect the nearest neighbors.  A $3 \times 3$~pixel subarray is considered:
\begin{equation}
s=\left( \begin{array}{ccc} s_0 & s_1 & s_2 \\ s_3 & s_4 & s_5 \\ s_6 &
s_7 & s_8 \end{array} \right).
\label{eqn:pixorder}
\end{equation}

The interpixel charge coupling can be represented by a detector point spread function (PSF) that, when convolved with the subarray, ``blurs'' the charge between pixels.  This kernel is represented as:
\begin{equation}
\text{PSF = }\left( \begin{array}{ccc} 0 & \alpha  & 0 \\ \alpha  & 1-4
\alpha  & \alpha  \\ 0 & \alpha  & 0 \end{array} \right),
\label{eqn:psf}
\end{equation}
where $\alpha$ is the coupling coefficient.  In other words, $\alpha$ is the fractional crosstalk between the central pixel and its nearest neighbors.  The ``true'' signal, $\tilde{s}$, in a pixel is therefore a linear combination of the measured signal in the central pixel of interest, $s_4$, and its neighbors,
\begin{equation}
\tilde{s}_4=\underset{i=0}{\overset{8}{\sum }}a_is_i,
\label{eqn:lincomb}
\end{equation}
where $a_i$ is some fraction of the signal in pixel $s_i$.  In fact, Equation~\ref{eqn:lincomb} represents the convolution of a kernel that undoes the blurring caused by the PSF.  In other words, $a$ must be the inverse PSF,
\begin{equation}
\overline{\text{PSF}}*\text{PSF}=\left( \begin{array}{ccc} 0 & 0 & 0 \\
0 & 1 & 0 \\ 0 & 0 & 0 \end{array} \right),\label{eqn:idkernel}
\end{equation}
where $*$ is the convolution operator.  For the PSF of Equation~\ref{eqn:psf}, the inverse PSF is,
\begin{equation}
\overline{\text{PSF}}=\frac{\alpha }{1-9 \alpha +18 \alpha ^2} \left(
\begin{array}{ccc} 2 \alpha  & -1+2 \alpha  & 2 \alpha  \\ -1+2 \alpha 
& -5+\frac{1}{\alpha }+2 \alpha  & -1+2 \alpha \\ 2 \alpha  & -1+2
\alpha  & 2 \alpha \end{array} \right).
\label{eqn:dpsf}
\end{equation}
By the associative property of convolution,
\begin{equation}
f*(g*h)=(f*g)*h,\label{eqn:assoc}
\end{equation}
Equation~\ref{eqn:lincomb} can be written as,
\begin{multline} 
\tilde{s}_4=\frac{1}{1+9 \alpha  (-1+2 \alpha )} \left(s_4+\alpha
\left(-s_1-s_3-5 s_4-s_5-s_7 \right.\right. \\ \left.\left.+2 \alpha
\left(s_0+s_1+s_2+s_3+s_4+s_5+s_6+s_7+s_8\right)\right)\right).
\label{eqn:truesig}
\end{multline}

The ``true'' variance of the pixel can be shown using the classical propagation of errors,
\begin{equation}
\sigma _{\tilde{s}_4}^2=\underset{j=0}{\overset{8}{\sum
}}\underset{i=0}{\overset{8}{\sum }}\frac{\partial \tilde{s}_4}{\partial
s_i}\frac{\partial \tilde{s}_4}{\partial s_j}C_{i,j},
\label{eqn:properr}
\end{equation}
where $C_{i,j}$ is the covariance between pixels $i$ and $j$.  The covariance matrix is given as
\begin{equation}
C_{ij} = \frac{1}{N-1}\sum_{n=1}^N (i_n-\overline{i})(j_n-\overline{j}),
\label{eqn:covariance}
\end{equation}
where $N$ is the number of pixels.  \citet{moo:06} show that to bring a one percent correlation signal up to the noise level, the number of samples must satisfy the equation
\begin{equation}
N\gg\frac{1}{\alpha^2}.
\label{eqn:nsamples}
\end{equation}
A coupling coefficient of $\sim$1\% requires $>10^6$~samples.

The covariance matrix can be found experimentally (see Section~\ref{sec:fullsym1}) with $>10^6$~independent integrations, but  our archival data set is limited to $<10$~integrations for each FPA.    To obtain the necessary number of samples in a realistic period of time requires a unique detector readout mode capable of sampling only small windows of pixels.  While such readout modes are currently being implemented for projects such as the James Webb Space Telescope (JWST), these readout modes are unavailable for HST because they are not required by the planned science.

Alternatively, the covariance matrix may be derived from theory under the assumption that noise in neighboring pixels is correlated solely with shot noise in the pixel of interest.  The theoretical covariance matrix is given as:
\begin{equation}
\scriptsize
C_{i,j} = \left(
\begin{array}{ccccccccc}
 \sigma ^2 & 0 & 0 & 0 & 0 & 0 & 0 & 0 & 0 \\
 0 & \sigma ^2 & 0 & 0 & \frac{\alpha  \sigma ^2}{1-4 \alpha } & 0 & 0 & 0 & 0 \\
 0 & 0 & \sigma ^2 & 0 & 0 & 0 & 0 & 0 & 0 \\
 0 & 0 & 0 & \sigma ^2 & \frac{\alpha  \sigma ^2}{1-4 \alpha } & 0 & 0 & 0 & 0 \\
 0 & \frac{\alpha  \sigma ^2}{1-4 \alpha } & 0 & \frac{\alpha  \sigma ^2}{1-4 \alpha } & \sigma ^2 & \frac{\alpha  \sigma ^2}{1-4 \alpha } & 0 & \frac{\alpha  \sigma ^2}{1-4 \alpha } & 0 \\
 0 & 0 & 0 & 0 & \frac{\alpha  \sigma ^2}{1-4 \alpha } & \sigma ^2 & 0 & 0 & 0 \\
 0 & 0 & 0 & 0 & 0 & 0 & \sigma ^2 & 0 & 0 \\
 0 & 0 & 0 & 0 & \frac{\alpha  \sigma ^2}{1-4 \alpha } & 0 & 0 & \sigma ^2 & 0 \\
 0 & 0 & 0 & 0 & 0 & 0 & 0 & 0 & \sigma ^2
\end{array}
\right).\label{eqn:covarmat}
\end{equation}
Equation \ref{eqn:covarmat} requires knowledge of only the coupling coefficient, $\alpha$, and the measured noise, $\sigma$.  We use Equation \ref{eqn:covarmat} throughout our analysis.

Substituting Equations~\ref{eqn:truesig} and~\ref{eqn:covarmat} into Equation~\ref{eqn:properr} gives the ``true'' variance in the pixel of interest,
\begin{equation}
{\sigma }_{\tilde{s}_4}^2 \approx \left(1+8 \alpha +52 \alpha ^2 + \dots \right) \sigma_{s_4}^2,
\label{eqn:newvar}
\end{equation}
where $\sigma_{s_4}^2$ is the measured variance.  \citet{moo:06} derived the same results to first order in $\alpha$ using Fourier methods.  \footnote{They did not explicitly write out higher order terms.}

\subsection{Validation}
\label{sec:validation}

Monte Carlo simulations validate the propagation of errors technique given by Equation~\ref{eqn:newvar}.  Each simulation models charge coupling in a uniformly illuminated $1024\times 1024$ detector array.  Illumination levels range in brightness from $10\leq s\leq 10^5~e^-~\text{pixel}^{-1}$, where $s$ is the mean integrated flux per pixel.  The shot noise is distributed with a Poisson distribution.  Charge coupling is simulated via convolution of a PSF given by Equation \ref{eqn:psf}, with a coupling coefficient $\alpha=1.5\%$.  A Gauss-normal distribution of read noise, $\sigma _{\text{read}}=15~e^-$, is added last.  We choose these parameters to be consistent with the actual data observed in Section \ref{sec:results}.

The ``true'' variance of the data when no charge coupling effects are present is given as:
\begin{equation}
\sigma_{\text{true}}^2 = s + \sigma_{\text{read}}^2
\label{eqn:true}
\end{equation}
The simulated charge coupling attenuates the variance.  We implement three different techniques to recover the ``true'' variance: (1) a 1+8$\alpha$ approximation, (2) the autocorrelation technique given by \citet{moo:06}, and (3) the propagation of errors technique described above in Section \ref{sec:fullsym}.

The first method attempts to restore the variance by multiplying by a $1+8\alpha$ coefficient.  This coefficient accounts for the $\sim1-8\alpha$ variance loss given by a small $\alpha$ approximation in equation 28 of \citet{moo:06}.  The technique is commonly used within the near-infrared community because of its straightforward implementation.  The coupling coefficient, $\alpha$, is typically measured during a dark exposure by utilizing the single pixel reset feature of the H2RG multiplexer to set a single pixel to a voltage different from its neighbors \citep{ses:08}.  The crosstalk induced by IPC on neighboring pixels is observed directly.

The autocorrelation procedure described by \citet{moo:06} relates the ``true'' variance to the autocorrelation matrix:
\begin{multline}
\sigma_{\rm true}^2 = \frac{1}{2N}(\sum_{i,j}D^2[i,j] + 2\sum_{i,j}D[i,j]D[i+1,j] + \\
\left. \left. \left. \left. \left. \left. \left. \left. 2\sum_{i,j}D[i,j]D[i,j+1]),\right.\right.\right.\right.\right.\right.\right.\right.
\label{eqn:moore34}
\end{multline}
where $D$ is the difference of two equally illuminated flat fields and $N$ is the number of pixels.  Equation \ref{eqn:moore34} makes a small $\alpha$ approximation and considers all contributions from pixels beyond the four nearest neighbors to be negligible.  In addition, \citet{moo:06} assume IPC to be the dominant coupling mechanism.  Other coupling mechanisms, however, typically are present, such as correlations in the the photon arrival (e.g. Bose=Einstein or Hanbury-Brown-Twiss) and carrier diffusion (e.g. carrier-carrier interaction).  Despite these assumptions, this variance estimator is typically favored by the near-infrared community \citep{bro:06,fin:06}.

 
Finally, method three implements the propagation of errors technique described in Section \ref{sec:fullsym}.  The ``true'' variance is given by Equation \ref{eqn:newvar}.  The covariance matrix in Equation \ref{eqn:covarmat} directly measures contributions from all additional coupling mechanisms that may be present.  Additional data could potentially resolve the different coupling mechanism contributions.  Nonetheless, like \citet{moo:06}, we assume IPC to be the only coupling mechanism for the purposes of this paper.  We simulate our real data by creating only 10 integrations.  Since the propagation of errors technique requires $10^6$~independent integrations to calculate the covariance matrix experimentally, we instead use the theoretical form given by Equation \ref{eqn:covarmat}, where the coupling coefficient, $\alpha$, and the noise, $\sigma$, are measured directly.  Section \ref{sec:alpha} describes the measurement of the coupling coefficient in more detail below.

\begin{figure} 
\begin{center}
\plotone{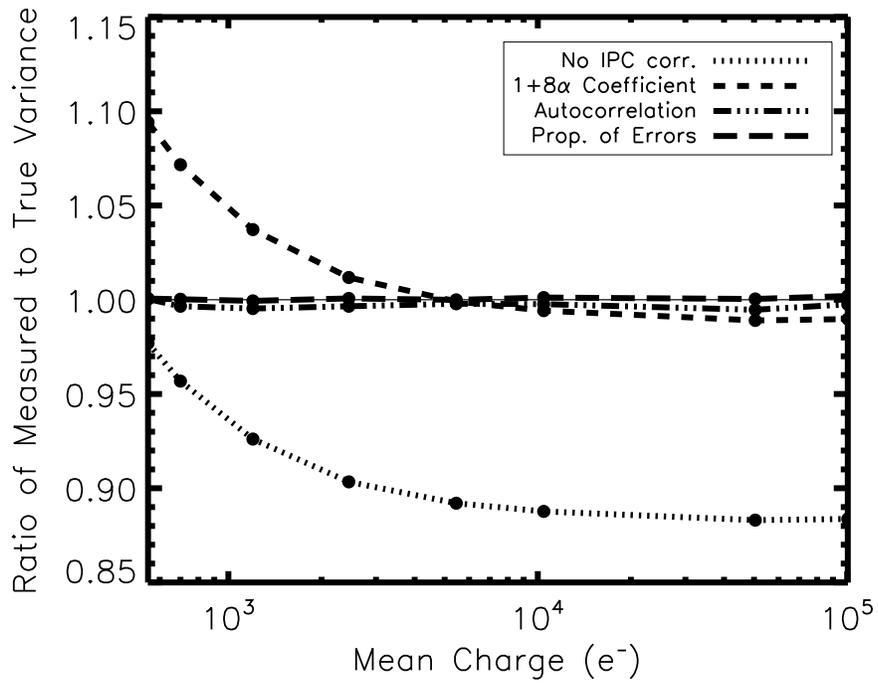}
\caption{Monte Carlo validation of Equation~\ref{eqn:newvar}. Of the models that were tested, the propagation of errors technique best recovers the ``true'' variance. The four models that were tested include: no correlated noise correction (``No IPC corr.''), multiplication by corrective coefficient (``$1+8\alpha$ Coefficient''),  implementation of equation 34 from \citet{moo:06} (``Autocorrelation''), and application of the correlated noise correction using propagation of errors (``Prop. of Errors'').}
\label{f4}
\end{center}
\end{figure}

Figure \ref{f4} plots the ratio of the measured to the true variance versus signal for each of the above methods.  As we predict, no correction results in an underestimated variance.  The rise at small signals is due to the dominating read noise component.  At high signal levels ($>500~e^-$), the $1+8\alpha$ approximation effectively estimates the variance, although there is a disagreement of $\sim$1\% due to the small $\alpha$ approximation.  At low signal levels, however, this technique is ineffective because the dominanting read noise does not couple between pixels.  The autocorrelation technique restores the ``true'' variance to within 1\% at all signal levels.  This method, however, does not fail at low signal levels because read noise does not contribute to the autocorrelation terms.

The propagation of errors technique is most effective at restoring the ``true'' variance at all signal levels. Again, there is no read noise contribution to the autocorrelation coefficients.  This technique is not as straightforward as the autocorrelation technique because it requires a direct measurement of the coupling coefficient, but it allows us to restore the variance most effectively.  For this reason, we implement the propagation of errors technique throughout the rest of this paper.

\subsection{Measuring the Coupling Coefficient, $\alpha$}
\label{sec:alpha}

The propagation of errors technique (Equation~\ref{eqn:newvar}) requires knowledge of the coupling coefficient, $\alpha$.  Unlike the simulations in Section \ref{sec:validation}, where the coefficient is a known input, the actual coupling coefficient must be measured directly from the data.  \citet{moo:06} show that $\alpha$ can be extracted directly from the autocorrelation matrix of uniformly illuminated exposures.  The ratio, $r$, of the nearest neighbor and unshifted autocorrelation coefficients is given as:
\begin{equation}
r = \frac{2\alpha(1-4\alpha)}{(1-4\alpha)^2 + 4\alpha^2}.
\label{eqn:ratio}
\end{equation}

Table \ref{tab1} lists the autocorrelation matrix of FPA 152.  The other detectors yield similar matrices.  The autocorrelation matrix is computed from the differenced image of two identically illuminated exposures.  To account for $1/f$-noise variations, we averaged differenced images computed from pairs of flat-field exposures with varying delay times between exposures.  We assume symmetry in both the horizontal and vertical directions so that the autocorrelation of each pixel with itself (i.e. unshifted) is given by the upper left ([0,0]) matrix value.

All values are normalized to the unshifted autocorrelation coefficient.  The nearest neighbors ([0,1] and [1,0]) have correlations that are typically between 1-5\%.  Even diagonal and second neighbor pixels show some correlation although the degree is much less than the nearest neighbors.  Solving for $\alpha$, Equation \ref{eqn:ratio} indicates the coupling coefficients are typically between 1-2\%.  Figure \ref{f5} plots the corrected conversion gain versus signal and compares it to the uncorrected values for FPA 152.  As predicted by the Monte Carlo simulations, the uncorrected gain is typically overestimated by $\sim13$\% for coupling coefficients of $\sim$1.5\%.

\subsection{Measuring the Conversion Gain, $g_c$}
\label{sec:gain}

\begin{figure}
\begin{center}
\subfigure[FPA 152]{
\label{f5a}
\plotone{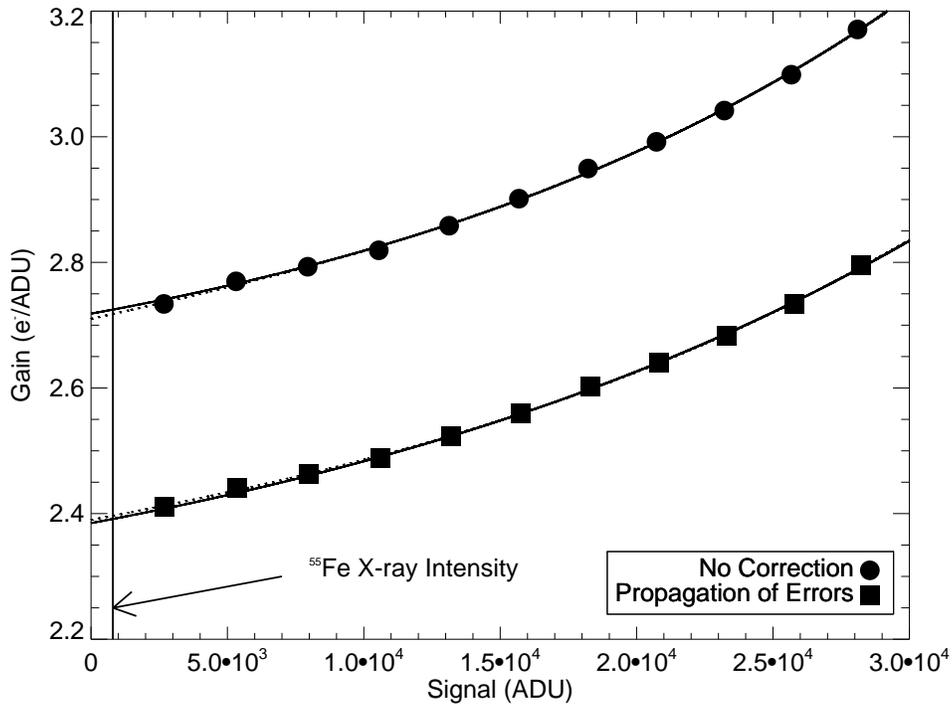}}\\
\subfigure[FPA 153]{
\label{f5b}
\plotone{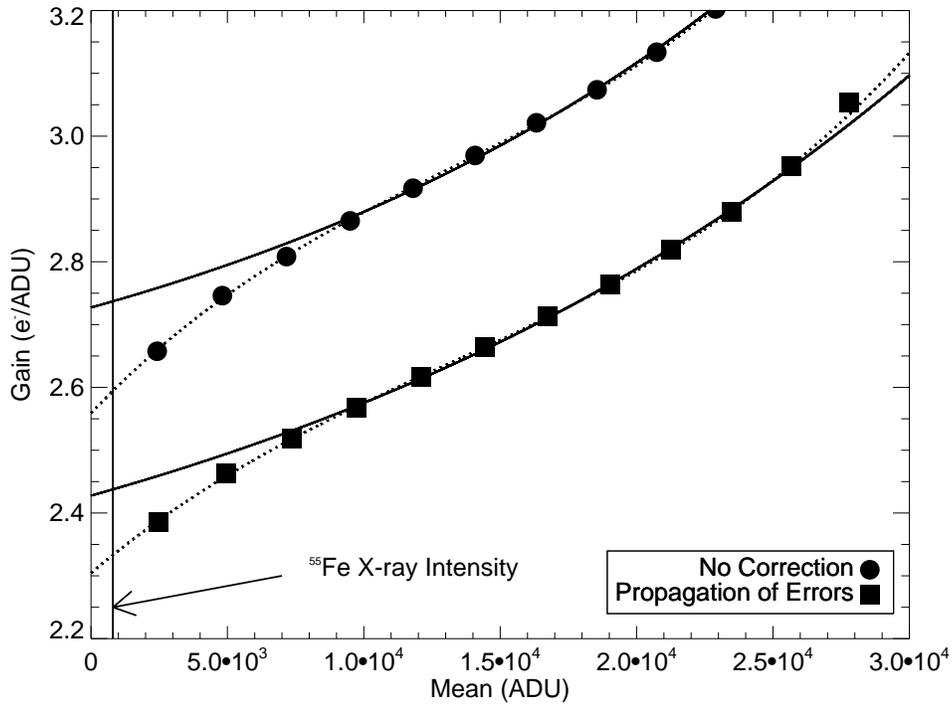}}
\caption{Gain ($e^-$/ADU) versus signal intensity (ADU) for FPAs 152 and 153.  The corrected gain has had the ``true'' variance restored via the propagation of errors.  As predicted by the Monte Carlo simulations, the uncorrected gain is typically overestimated by $\sim13$\%.  The relevant gain is measured at signal levels consistent with the \Fe X-ray intensity, which are typically $\sim$800-1000 ADU, as indicated by the vertical lines.  We measure this number by extrapolating a best fitting line down to lower signal levels.  While both the polynomial (dotted line) and exponential (solid line) fits yield similar results in the case of FPA 152 (a), the polynomial fit appears to be the most appropriate fit in cases where the gain rapidly declines at lower signal levels, as in the case of FPA 153 (b).  (FPA 153 is the worst case within our data set.)  We therefore prefer the polynomial fit for our extrapolation in all cases.}
\label{f5}
\end{center}
\end{figure}

The relevant conversion gain, $g_c$, should correspond to signal levels equivalent to the \Fe X-ray intensity, which is highlighted by the vertical line in Figure \ref{f5} ($\sim$800 ADU).  The figure reveals, however, that the signal levels of the gain data are typically higher.  While an obvious solution would be to retake the necessary data, our analysis is limited to archival data.  Instead, we attempt to estimate the conversion gain at low signal levels and minimize the associated systematic error.

Figure \ref{f5} shows a correlation between the conversion gain and signal level.  This trend arises from the relationship between the capacitance of the diode and bias voltage, which is well understood for HAWAII devices \citep[e.g.][]{fin:06}.    The capacitance increases with signal level because the voltage, and hence the depletion width of the pn junction, across the diode decreases.  For low signal levels (i.e. small capacitances) an electron generates a larger voltage swing and the gain (in $e^-$/ADU) is smaller.

To better understand the gain behavior of these devices at low signal levels, we performed alternate gain measurement techniques, including an analysis of the photon transfer of warm pixels in dark ramps and solving for the gain in dark ramps via equation 1 of \citet{rau:07}.  We also compared our results to ultra-low light level data collected for similar FPAs undergoing testing for the JWST Near-Infrared Spectrograph project (Rauscher, priv. comm.).  These results all confirm the gain decreases with signal level and that the decrease is smooth and continuous.

We therefore choose to fit a curve to the high signal gain data and extrapolate to lower signal levels.  The choice of curve to fit is unclear because the true behavior of the gain is not well described at these low signal levels, where the data tend to be noisier.   Figure \ref{f5a} suggests that an exponential fit is sufficient.  Figure \ref{f5b}, however, indicates that a polynomial provides a better fit.  We find the likely gain is bound by the extremes of these two curves and take the conversion gain to be an average of the two.  Table \ref{tab2} lists the ``true'' gain for all the detectors.  We discuss the systematic error associated with this method in Section \ref{sec:syserr}.

\section{Results}
\label{sec:results}

\subsection{Pair-Production Energy}
\label{sec:ppe}

\begin{figure} [h]
\begin{center}
\epsscale{1}
\plotone{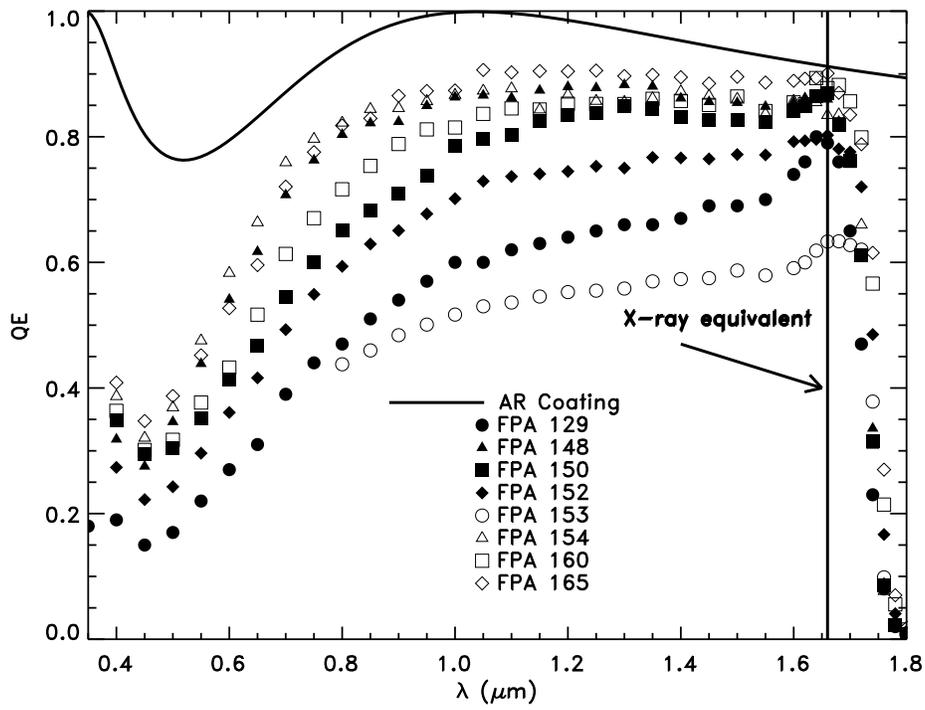}
\caption{Measured quantum efficiency (QE) of each detector, as well as the reflective coating transmission model, as a function of wavelength.  The vertical line denotes the photon wavelength corresponding to the equivalent penetration depth of the \Fe X-rays.  Like these long wavelength photons, the X-rays typically penetrate deep into the detector.  We take the equivalent QE at longer wavelengths just before the sharp roll-off, at which the electron capture efficiency no longer dominates the QE measurements.}
\label{f6}
\end{center}
\end{figure}

We calculate the pair-production energy by dividing the X-ray response (in $e^-$) into the energy of a single \Fe X-ray, which includes a weighted contribution from the blended K$\beta$ peak.  The X-ray response (in $e^-$) is given by the product of the measured \Fe peak ADU intensity (measured in Section \ref{sec:methods}) with the ``true'' conversion gain (measured in Section \ref{sec:math}).  We must first divide the gain by the detector internal QE to account for charge that may have recombined before being collected in the pixels' depletion regions.  Figure \ref{f6} plots the measured QE of each detector.

The relevant QE measurement from Figure \ref{f6} depends on the wavelength of the photons with similar penetration depths as \Fe X-rays.  The X-ray penetration in HgCdTe follows an exponentially falling probability distribution.  The attenuation length is defined as the depth, $x$, into the material, measured along the surface normal, where the intensity of X-rays falls to 1/$e$ of its value at the surface.  Our model, 
\begin{equation}
x = 2.025 - 9.472\times10^{-4}e_\gamma + 1.8\times10^{-7}e_\gamma^2,
\end{equation}
is based on data from \citet{hen:93} and subsequent modeling\footnote{http://henke.lbl.gov/optical\_constants/atten2.html}, where $x$ is given in microns and $e_\gamma$ is the X-ray energy in eV.  The \Fe X-ray penetration depth is towards the very back of these detectors and has an associated cutoff wavelength of $\sim$1.68 \micron~\citep{scha:01}.

The QE of a detector, however, typically combines the effects of several external factors, such as the number of photons that pass through the detector and the number of photons reflected by the detector's anti-reflective (AR) coating, and the internal QE, which includes only losses in the detector crystal.  Since our tests measure the detector response to individual X-rays, which are either absorbed or not, we are concerned only with the internal QE.  For the plot in Figure \ref{f6}, we subtract the AR coating contribution with a model AR curve.  We also assume that all photons are absorbed as long as they are short-ward of the roll-off at long wavelengths.  Therefore, instead of measuring the QE at precisely the associated cutoff wavelength calculated above ($\sim$1.68 \micron), we measure the QE just before the sharp roll-off, as indicated by the vertical line.  We discuss the systematic error associated with this method in Section \ref{sec:syserr}. 

\begin{figure} [h]
\begin{center}
\plotone{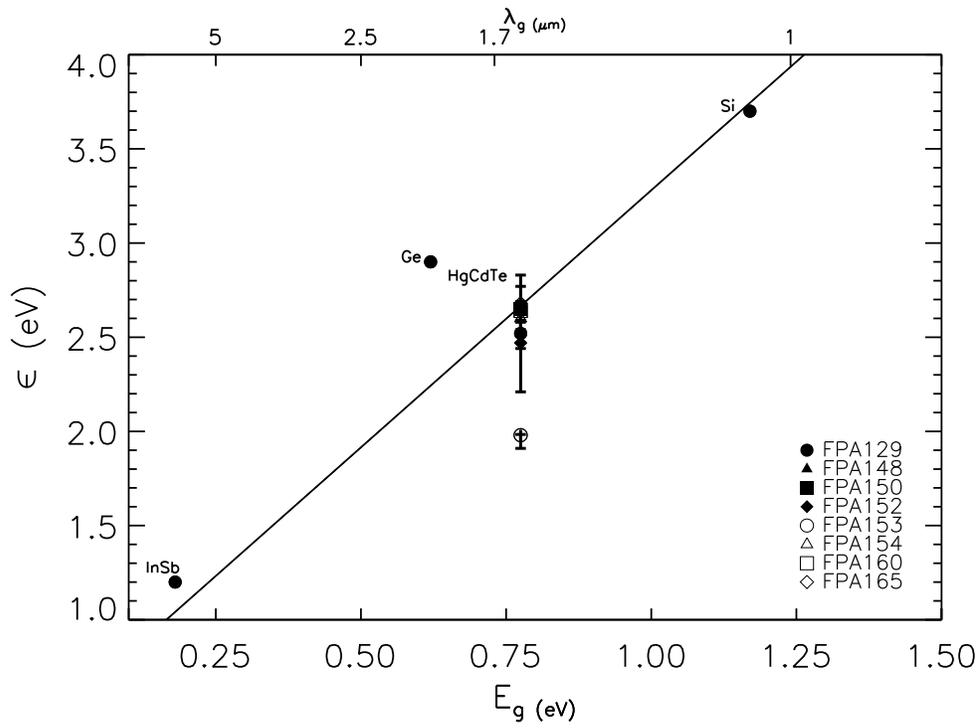}
\caption{Measured pair creation energies, $\epsilon$, of the HST WFC3 1.7 \micron~HgCdTe detectors plotted against the semiconductor bandgap energy, $E_g$, and the cutoff wavelength corresponding to $E_g$, $\lambda _{\text{g}}=hcE_g^{-1}$.  Also plotted in this figure are several other semiconductor materials published in \citet{ali:75}'s figure 1.  The solid line is the semi-empirical relationship between $\epsilon$ and $E_g$, derived in \citet{ali:75}.}
\label{f7}
\end{center}
\end{figure}

Table \ref{tab2} lists the resulting pair-production energy for each detector.  Figure \ref{f7} plots the pair-production energy versus cutoff wavelength and bandgap energy.  FPAs 129 and 153 stand out from the rest.  Interestingly, these 2 detectors exhibit traits and data sets uncharacteristic of the typical FPAs that we tested.  

For example, FPA 129 has a large error bar traceable to the unique data available for this FPA.  Unlike the other FPAs, the conversion gain data for FPA 129 consist of only 5 data points, all of which have signal levels $>$5000 ADU.  Figure \ref{f5b} indicates that the slope of the conversion gain only begins to change at signal levels $<$5000 ADU.  Considering the \Fe intensity is $\sim$800 ADU, we approximate the lower signal gain from an analysis of the photon transfer of warm pixels in dark ramps, which has a large associated error.  The resulting extrapolation of the FPA 129 data set is significantly less constrained and, subsequently, has a much larger systematic error than the other FPAs.  Given the limited data set, we do not believe that the error associated with FPA 129 represents the typical systematic error associated with our procedure and therefore exclude this data set from our final average.

The data point for FPA 153 in Figure \ref{f7} is clearly isolated from the other FPAs.  FPA 153 has a significantly lower QE than the other FPAs, especially at the cutoff wavelength.  In fact, we find that the QE for this FPA tends to swing $\pm20\%$ between different data sets.  Furthermore, FPA 153 shows the most rapid decline in conversion gain at low signal levels.  We do not completely understand the odd behavior FPA 153 exhibits, but we recognize that this is not standard behavior for these FPAs.  We therefore do not include the measurement in our final average.

Averaging over the remaining FPAs, the pair-production energy of 1.7 \micron~HgCdTe is 2.61~$\pm$~0.16 eV, which corresponds to 2273~$\pm$~137 electrons produced by each \Fe X-ray.  For comparison purposes, Figure \ref{f7} plots the pair-production energy of several other common semiconductor materials, as well as the semi-empirical trend given by \citet{ali:75}.  The average pair-production energy for 1.7 \micron~HgCdTe falls within a few percent of the value predicted by \citet{ali:75}, which is within the given error.  We did not necessarily predict this agreement given that the trend-line is only semi-empirical.  In fact, the measured pair-production energy for germanium does not fall even close to the trend-line.  Nonetheless, Figure 1 of \citet{ali:75} reveals that a majority of materials do fall reasonably close to the line.


The overall consistency of the 1.7 \micron~results indicates that the \Fe results may be an easy, straightforward sanity check for a given FPA.  For example, the inconsistent results for FPAs 129 and 153 indicated a bad data set and a misbehaving FPA, respectively.  In other words, if the calculated pair-production energy for a given FPA falls far from the predicted value, the testers may wish to review the other FPA characteristics.

\subsection{Statistical Errors}
\label{sec:staterr}

We identify four possible statistical errors in our measurements: the (1) detector response to the X-rays in ADU, (2) the QE, (3) the conversion gain, and (4) the coupling coefficient, $\alpha$.  Given the large number of samples in all cases, the associated statistical errors are negligible.  Specifically, we determine the detector response from the \Fe peak in Figure \ref{f2}.  In this case, the large number of pixels greatly reduces the error of the mean to $<1$\%.  Similarly, the large number of pixels reduces the error bars of the measured QE and conversion gain to $<1$\%.  The curves fit to the gain data subsequently have negligible error bars.  Finally, Equation \ref{eqn:covariance} indicates that a coupling coefficient of $\sim$1\% requires $>10^6$~samples to obtain a signal-to-noise $>10$.  In terms of our autocorrelation matrices, individual exposures $>10^6$~pixel samples (1024$\times$1024 array) easily satisfy this requirement.

\subsection{Systematic Errors}
\label{sec:syserr}
The systematic errors tend to be larger than the statistical errors described above.  Specifically, we identify three possible systematic errors in our methods: measurements of the (1) detector response to the X-rays in ADU, (2) the QE, and (3) the conversion gain.  We measure the detector response peak by fitting the histogram in Figure \ref{f2b} with a combined gaussian and polynomial using IDL's GAUSSIAN routine.  The peak of the distribution varies, however, depending on the signal threshold chosen to indicate a well-centered hit in Figure \ref{f2a}.  We set the threshold at the ``kink'' where the low-intensity plateau and gaussian peak meet.  This point is rather arbitrary and may not be optimal.  Varying the threshold by a reasonable amount (e.g. $<$10 ADU) in either direction shifts the gaussian peak in Figure \ref{f2b} by a similar amount.  This shift, which typically results in only $\sim$1\% error, presents only a minor contribution to the overall systematic error.

Measuring the QE introduces an additional systematic error.  As discussed in Section \ref{sec:ppe}, the \Fe X-rays have an associated cut-off wavelength of $\sim$1.68 \micron.   At these long wavelengths, however, the QE is a combination of both internal losses and unabsorbed photons.  Concerned only about electrons generated by absorbed X-rays, we desire a QE that measures internal losses only.   We assume that all photons are absorbed short-ward of the QE roll-off at long wavelengths in Figure \ref{f6}.  Instead of measuring the QE at 1.68 \micron, we estimate a shorter wavelength at which the QE represents internal losses only, as indicated by the vertical line in Figure \ref{f6}.  We assign an error of 0.05 \micron~to this estimated wavelength as it is not entirely clear where the roll-off begins.  This error corresponds to a systematic error for the QE, which depends on the degree of slope in the QE curve.  For all FPAs, the resulting QE systematic error is $<$5\%.

The measurement of the conversion gain introduces the largest systematic error to our results.  Figure \ref{f5} shows the measured gain as a function of signal level.  The desired gain for our purposes must be measured at signal levels comparable to the X-ray intensities, which are typically $\sim$800 ADU.  Our archival gain data, however, is limited to higher signal levels than necessary.  As discussed in Section \ref{sec:gain}, we therefore choose to fit a curve to the high signal gain data and extrapolate to lower signal levels.  Given the contrasting nature of the trends in Figures \ref{f5a} and \ref{f5b}, we find the likely conversion gain at low signal levels is bound by the extremes of both an exponential and polynomial fit.  We therefore take the conversion gain to be an average of these two fits.  The associated systematic error is given by the difference between the average and the values of the two curves at the desired intensity (in this case, $\sim$800 ADU).  We find a systematic error of $<$10\% in all cases.

The systematic errors are listed in Table \ref{tab2} and included on the plot in Figure \ref{f7}.  While these systematic errors contribute to no more than $\sim$11\% in quadrature, we believe these errors can be significantly reduced with more appropriate data.  For example, we measure the gain by extrapolating our data sets down to low signal levels.  Future tests should be sure to measure the gain directly at these signal levels.  Such foresight would reduce the systematic errors by at least 50\%, if not more.

\
\section{Summary
\label{sec:summary}}

In this paper, we have calibrated the \Fe X-ray response of Teledyne 1.7 \micron~HgCdTe flight grade detectors for the HST WFC3.  Our conversion of the X-ray intensities from ADU into electrons implements a technique that restores the ``true'' gain of the detectors using the classical propagation of errors.  The advantage to this technique is that it allows for a direct measure of the covariance matrix from the data.  Monte Carlo simulations have validated this technique as being most effective at restoring variance that is lost due to charge coupling effects.

The sample of eight 1.7 \micron~H1R detectors yields 2273~$\pm$~137 electrons per \Fe X-ray, which corresponds to a pair-production energy for 1.7 \micron~HgCdTe of 2.61~$\pm$~0.16 eV.  This result includes both statistical and systematic errors.  Better designed tests can significantly reduce these errors in the future.  Nonetheless, these results now make the \Fe gain measurement technique of potential interest for near-infrared arrays, as it has been in the CCD community for many years.

The \Fe X-ray technique provides a valid measurement of the conversion gain at signal levels comparable to the \Fe X-ray intensity.  This low signal level ($\sim$800 ADU) is consistent with the signal levels at which the gain is typically reported for WFC3 detectors.  This technique provides a straightforward alternative to the photon transfer technique, which can overestimate the conversion gain within HgCdTe detectors by up to $\sim$20\% (and Si-PIN arrays by more than a factor of 2) when not accounting for charge coupling effects, such as IPC.  The known detector response to \Fe also will aid future detector design and performance studies, such as measuring drifts in the system gain or the internal QE.

Although these measurements were made using Teledyne HAWAII detectors, we believe that they are likely to be applicable to HgCdTe detectors from other vendors, so long as differences in cutoff wavelength are correctly accounted for.  Users of Teledyne's older substrate-on should not use this method as the substrate absorbs the X-rays.  Already, \citet{fin:08} have performed similar measurements of the  \Fe X-ray response for 2.5 \micron~HgCdTe H2RG detectors.  They report results consistent with the overall trend in Figure \ref{f7}, but these results fall significantly above the predicted line compared to the 1.7 \micron~results.  The authors use different calibration techniques and do not report a detailed review of their systematic errors.  In the future, we plan to implement our method to check the 2.5 \micron~results, as well as test HgCdTe with other cutoff wavelengths (e.g. 5 \micron).

\
\

This research was supported by NASA as part of the HST's Wide Field Camera 3 program and as part of the James Webb Space Telescope Project.  Ori Fox wishes to thank NASA's Graduate Student Researcher Program for a grant to UVa.

\begin{appendix}
\label{sec:appdx}

\section{How to Proceed Without Assuming the form of the Covariance
Matrix as a Prior}

\subsection{Correlated Noise \& Partially Symmetric PSF
\label{sec:partsym}}

In most HST detector arrays, the detector PSF is symmetric in
reflection about the $x$ and $y$ axes to within a few a few percent. For
the most accurate measurements of conversion gain, this asymmetry needs
to be accounted for when it is detectable. In this case, the detector
PSF can be written as follows,
\begin{equation}
\text{PSF}=\left( \begin{array}{ccc} 0 & \alpha  & 0 \\ \beta  & 1-2
(\alpha +\beta ) & \beta  \\ 0 & \alpha  & 0 \end{array}
\right),\label{eqn:partsympsf}
\end{equation}
and the deconvolution kernel is,
\begin{multline}
\overline{\text{PSF}}=\frac{\alpha  \beta }{(-1+3 \alpha ) (-1+3 \beta )
(-1+3 \alpha +3 \beta )}\\ \times \left(
\begin{array}{ccc}
 -2+3 \alpha +3 \beta  & -2+3 \alpha +\frac{1-3 \alpha }{\beta }+3 \beta  & -2+3 \alpha +3 \beta  \\
 -2+3 \alpha +\frac{1-3 \beta }{\alpha }+3 \beta  & \frac{-1+3 \alpha ^2 (-1+\beta )+(4-3 \beta ) \beta +\alpha  (4+\beta  (-11+3 \beta ))}{\alpha  \beta } & -2+3 \alpha +\frac{1-3 \beta }{\alpha }+3 \beta  \\
 -2+3 \alpha +3 \beta  & -2+3 \alpha +\frac{1-3 \alpha }{\beta }+3 \beta  & -2+3 \alpha +3 \beta 
\end{array}
\right)
\end{multline}

Following the same methods as were used in Section~\ref{sec:fullsym}, we
arrive at an expression for $\tilde{\sigma }_4^2$ that is accurate to
second order in $\alpha$ and $\beta$,
\begin{multline}
\tilde{\sigma }_4^2=C_{4,4}+\alpha \left(-C_{1,4}-C_{4,1}+4
C_{4,4}-C_{4,7}-C_{7,4}\right)\\
+\beta  \left(-C_{3,4}-C_{4,3}+4 C_{4,4}-C_{4,5}-C_{5,4}\right)\\
+\alpha ^2 \left(C_{1,1}-5 C_{1,4}+C_{1,7}-5 C_{4,1}+16 C_{4,4}-5
C_{4,7}+C_{7,1}-5 C_{7,4}+C_{7,7}\right)\\
+\beta ^2 \left(C_{3,3}-5 C_{3,4}+C_{3,5}-5 C_{4,3}+16 C_{4,4}-5
C_{4,5}+C_{5,3}-5 C_{5,4}+C_{5,5}\right)\\
+\alpha  \beta  \left(2 C_{0,4}+C_{1,3}-6 C_{1,4}+C_{1,5}+2
C_{2,4}+C_{3,1}-6 C_{3,4}+C_{3,7}+2 C_{4,0}-6 C_{4,1}+2 C_{4,2} \right.
\\ \left. -6 C_{4,3}+24 C_{4,4}-6 C_{4,5}+2 C_{4,6}-6 C_{4,7}+2
C_{4,8}+C_{5,1}-6 C_{5,4}+C_{5,7}+2 C_{6,4}\right. \\ \left. +C_{7,3}-6
C_{7,4}+C_{7,5}+2
C_{8,4}\right)
\label{eqn:partsymvar}
\end{multline}
Because the mathematics are complex, we do not show the steps here,
other than to note that all computations were done using the {\em
Mathematica} symbolic mathematics program.

\subsection{Correlated Noise \& Completely Symmetric PSF
\label{sec:fullsym1}}

For our calibration of \Fe X-rays, we believe that the method described
in this paper (with extensions to remove the need to assume a functional
form for the detector PSF as a prior) is likely to have the advantage
that the covariance matrix can be explicitly computed from the data to
serve as a diagnostic of other correlations that may be present in the
data in addition to IPC. Unfortunately, doing this requires specialized
detector readout modes that have not been implemented for HST because they are not required by the planned science.

If we do not assume a theoretical form for the covariance matrix,
Equation~\ref{eqn:newvar} can be written as follows,
\begin{multline}
\tilde{\sigma }_4^2\approx C_{4,4}+\alpha 
\left(-C_{1,4}-C_{3,4}-C_{4,1}-C_{4,3}+8
C_{4,4}-C_{4,5}-C_{4,7}-C_{5,4}-C_{7,4}\right)\\ +\alpha ^2 \left(2
C_{0,4}+C_{1,1}+C_{1,3}-11 C_{1,4}+C_{1,5}+C_{1,7}+2
C_{2,4}+C_{3,1}+C_{3,3}-11 C_{3,4}+C_{3,5} \right.\\\left. +C_{3,7}+2 C_{4,0}-11
C_{4,1}+2 C_{4,2}-11 C_{4,3}+56 C_{4,4}-11 C_{4,5}+2 C_{4,6}-11
C_{4,7}+2 C_{4,8}\right.\\\left.+C_{5,1}+C_{5,3}-11 C_{5,4}+C_{5,5}+C_{5,7}+2
C_{6,4}+C_{7,1}+C_{7,3}-11 C_{7,4}+C_{7,5}+C_{7,7}+2
C_{8,4}\right) + \dots .\label{eqn:gensymvar}
\end{multline}
The individual $C_{i,j}$ can then be measured using the relation,
\begin{equation}
C_{i,j}=\frac{1}{n-1}\sum _{k=0}^{n-1}
\left(s_i-\overline{s_i}\right)_n\left(s_j-\overline{s_j}\right)_n.
\label{eqn:covar}
\end{equation}
In Equation~\ref{eqn:covar}, $n$ refers to the exposure sequence number,
which needs to be large for accurate measurement.

\end{appendix}

\clearpage
\begin{deluxetable}{ c c c c}
\tablewidth{0pt}
\tablecaption{Autocorrelation Matrix for FPA 152 \label{tab1}}
\tablecolumns{4}
\tablehead{}
\startdata
      1.000 & 0.041 & 0.006 & 0.005\\
      0.021 & 0.005 & 0.001 & 0.001\\
      1.4e-4 &  7.5e-4 &  2.6e-4 & -8.0e-6\\
      -3.5e-4 &   0.001 & 7.1e-4 & -4.0e-4
\enddata
\end{deluxetable}

\begin{deluxetable}{ c c c c c c c}
\tablewidth{0pt}
\tabletypesize{\tiny}
\tablecaption{Summary of Detector Properties \label{tab2}}
\tablecolumns{7}
\tablehead{
\colhead{\multirow{2}{*}{Serial Number}} & \colhead{\multirow{2}{*}{$\sigma_{\rm read}$}} & \colhead{Gain} & \colhead{Crosstalk} & \colhead{Gain} & \colhead{\multirow{2}{*}{Gain Correction}} & \colhead{\multirow{2}{*}{pair-production Energy}}\\
\colhead{} & \colhead{} & \colhead{(before correction)} & \colhead{(Hor/Vert)} & \colhead{(after correction)} & \colhead{} & \colhead{}\\
\colhead{} & \colhead{$e^-$~rms} & \colhead{$e^-$/ADU} & \colhead{\%} & \colhead{$e^-$/ADU} & \colhead{\%} & \colhead{eV}
}
\startdata
FPA 129 & 13.74 & 3.25 & 2.98/1.50 & 2.55 & 21.6 & 2.52 (0.012 $\pm$ 0.31)\\
FPA 148 & 14.59 & 2.64 & 1.93/1.17 & 2.34 & 11.6 & 2.64 (0.002 $\pm$ 0.07)\\
FPA 150 & 19.80 & 2.73 & 2.07/1.29 & 2.38 & 12.7 & 2.65 (0.002 $\pm$ 0.10)\\
FPA 152 & 19.04 & 2.72 & 1.93/1.12 & 2.40 & 11.9 & 2.47 (0.005 $\pm$ 0.03)\\
FPA 153 & 18.02 & 2.67 & 2.02/1.11 & 2.35 & 11.9 & 1.98 (0.005 $\pm$ 0.09)\\
FPA 154 & 18.38 & 2.67 & 1.97/1.25 & 2.23 & 16.6 & 2.60 (0.003 $\pm$ 0.05)\\
FPA 160 & 15.68 & 2.67 & 2.04/1.28 & 2.30 & 14.0 & 2.64 (0.002 $\pm$ 0.04)\\
FPA 165 & 13.37 & 2.62 & 2.22/1.46 & 2.18 & 17.0 & 2.68 (0.007 $\pm$ 0.06)
\enddata
\end{deluxetable}

\clearpage

\clearpage

\clearpage

\clearpage

\clearpage

\clearpage

\clearpage


\clearpage
\begin{thebibliography}{15}
\expandafter\ifx\csname natexlab\endcsname\relax\def\natexlab#1{#1}\fi

\bibitem[{Alig \& Bloom(1975)}]{ali:75}
Alig, R.~C., \& Bloom, S. 1975, Physical Review Letters, 35, 1522

\bibitem[{{Bai} {et~al.}(2007){Bai}, {Farris}, {Petersen}, \&
  {Beletic}}]{bai:07}
{Bai}, Y., {Farris}, M.~C., {Petersen}, A.~K., \& {Beletic}, J.~W. 2007, Proc. SPIE, 6690, 669004

\bibitem[{Brown {et~al.}(2006)Brown, Schubnell, \& Tarl{\'e}}]{bro:06}
Brown, M., Schubnell, M., \& Tarl{\'e}, G. 2006, \pasp, 118, 1443
  
\bibitem[{{Dorn} {et~al.}(2006){Dorn}, {Eschbaumer}, {Finger}, {Mehrgan}, {Meyer}, \& {Stegmeier}}]{dor:06}
{Dorn}, R.~J., {Eschbaumer}, S., {Finger}, G., {Mehrgan}, L., {Meyer}, M., \& {Stegmeier}, J. 2006, Proc. SPIE, 6276, 627607

\bibitem[{{Figer} {et~al.}(2004){Figer}, {Rauscher}, {Regan}, {Morse},
  {Balleza}, {Bergeron}, \& {Stockman}}]{fig:04}
{Figer}, D.~F., {Rauscher}, B.~J., {Regan}, M.~W., {Morse}, E., {Balleza}, J.,
  {Bergeron}, L., \& {Stockman}, H.~S. 2004, Proc. SPIE, 5167, 270-301 
  
\bibitem[{Finger {et~al.}(2006)Finger, Beletic, Dorn, Meyer, Mehrgan, Moorwood,
  \& Stegmeier}]{fin:06}
Finger, G., Beletic, J.~W., Dorn, R., Meyer, M., Mehrgan, L., Moorwood, A.
  F.~M., \& Stegmeier, J. 2006, Exp Astron, 19, 135
  
  \bibitem[{{Finger} {et~al.}(2006){Finger}, {Dorn}, {Meyer}, {Mehrgan},
  {Moorwood}, \& {Stegmeier}}]{fin:06}
   {Finger}, G., {Dorn}, R., {Meyer}, M., {Mehrgan}, L., {Moorwood}, A.~F.~M., \& {Stegmeier}, J. 2006, Proc. SPIE, 6276, 62760F
   
     \bibitem[{{Finger} {et~al.}(2008){Finger}, {Dorn}, {Eschbaumer}, {Hall},
  {Mehrgan}, {Meyer}, \& {Stegmeier}}]{fin:08} {Finger}, G., {Dorn}, R.~J., {Eschbaumer}, S., {Hall}, D.~N.~B., {Mehrgan}, L., {Meyer}, M., \& {Stegmeier}, J. 2008, Proc. SPIE, 7021, 70210P-70210P-13
 
\bibitem[{Fraser {et~al.}(1994)Fraser, Abbey, Holland, McCarthy, Owens, \&
  Wells}]{fra:94}
Fraser, G.~W., Abbey, A.~F., Holland, A., McCarthy, K., Owens, A., \& Wells, A.
  1994, Nuclear Instruments and Methods in Physics Research Section A, 350, 368

\bibitem[{Henke {et~al.}(1993)Henke, Gullikson, \& Davis}]{hen:93}
Henke, B.~L., Gullikson, E.~M., \& Davis, J.~C. 1993, Atomic Data and Nuclear
  Data Tables, 54, 181

\bibitem[{Janesick {et~al.}(1987)Janesick, Klaasen, \& Elliott}]{jan:87}
Janesick, J., Klaasen, K., \& Elliott, T. 1987, Optical Engineering, 26, 972

\bibitem[{Kavadias {et~al.}(1994)Kavadias, Misiakos, \& Loukas}]{kav:94}
Kavadias, S., Misiakos, K., \& Loukas, D. 1994, IEEE Transactions on Nuclear
  Science, 41, 397

\bibitem[{Moore {et~al.}(2006)Moore, Ninkov, \& Forrest}]{moo:06}
Moore, A.~C., Ninkov, Z., \& Forrest, W.~J. 2006, Optical Engineering, 45, 6402

\bibitem[{{Rauscher} {et~al.}(2007){Rauscher}, {Fox}, {Ferruit}, {Hill},
  {Waczynski}, {Wen}, {Xia-Serafino}, {Mott}, {Alexander}, {Brambora}, {Derro},
  {Engler}, {Garrison}, {Johnson}, {Manthripragada}, {Marsh}, {Marshall},
  {Martineau}, {Shakoorzadeh}, {Wilson}, {Roher}, {Smith}, {Cabelli},
  {Garnett}, {Loose}, {Wong-Anglin}, {Zandian}, {Cheng}, {Ellis}, {Howe},
  {Jurado}, {Lee}, {Nieznanski}, {Wallis}, {York}, {Regan}, {Hall}, {Hodapp},
  {B{\"o}ker}, {De Marchi}, {Jakobsen}, \& {Strada}}]{rau:07}
{Rauscher}, B.~J., {Fox}, O., {Ferruit}, P., {Hill}, R.~J., {Waczynski}, A.,
  {Wen}, Y., {Xia-Serafino}, W., {Mott}, B., {Alexander}, D., {Brambora},
  C.~K., {Derro}, R., {Engler}, C., {Garrison}, M.~B., {Johnson}, T.,
  {Manthripragada}, S.~S., {Marsh}, J.~M., {Marshall}, C., {Martineau}, R.~J.,
  {Shakoorzadeh}, K.~B., {Wilson}, D., {Roher}, W.~D., {Smith}, M., {Cabelli},
  C., {Garnett}, J., {Loose}, M., {Wong-Anglin}, S., {Zandian}, M., {Cheng},
  E., {Ellis}, T., {Howe}, B., {Jurado}, M., {Lee}, G., {Nieznanski}, J.,
  {Wallis}, P., {York}, J., {Regan}, M.~W., {Hall}, D.~N.~B., {Hodapp}, K.~W.,
  {B{\"o}ker}, T., {De Marchi}, G., {Jakobsen}, P., \& {Strada}, P. 2007,
  \pasp, 119, 768

\bibitem[{{Reach} {et~al.}(2005){Reach}, {Megeath}, {Cohen}, {Hora}, {Carey},
  {Surace}, {Willner}, {Barmby}, {Wilson}, {Glaccum}, {Lowrance}, {Marengo}, \&
  {Fazio}}]{reach:05}
{Reach}, W.~T., {Megeath}, S.~T., {Cohen}, M., {Hora}, J., {Carey}, S.,
  {Surace}, J., {Willner}, S.~P., {Barmby}, P., {Wilson}, G., {Glaccum}, W.,
  {Lowrance}, P., {Marengo}, M., \& {Fazio}, G.~G. 2005, \pasp, 117, 978

\bibitem[{{Schacham} \& {Finkman}(1985)}]{scha:01}
{Schacham}, S.~E., \& {Finkman}, E. 1985, Journal of Applied Physics, 57, 2001

\bibitem[{Schubnell {et~al.}(2006) {Schubnell}, {Barron}, {Bebek}, {Borysow}, {Brown},
  {Cole}, {Figer}, {Lorenzon}, {Bower}, {Mostek}, {Mufson}, {Seshadri}, {Smith}, \&
  {Tarl{\'e}}}]{sch:06} Schubnell, M., Barron, N., Bebek, C., Borysow, M., Brown, M.~G., Cole, D.,
  Figer, D., Lorenzon, W., Bower, C., Mostek, N., Mufson, S., Seshadri, S.,
  Smith, R., \& Tarl{\'e}, G. 2006, Proc. SPIE, 6276, 62760Q

\bibitem[{Scofield(1974)}]{sco:74}
Scofield, J.~H. 1974, Physical Review A, 9, 1041

\bibitem[{{Seshadri} {et~al.}(2008){Seshadri}, {Cole}, {Hancock}, \& {Smith}}]{ses:08}
{Seshadri}, S., {Cole}, D.~M., {Hancock}, B.~R., \& {Smith}, R.~M. 2008, Proc. SPIE, 7021, 702104

\end{thebibliography}
\end{document}